\documentclass[doublecol,linenumbers]{epl2} 

\usepackage{dsfont}
\usepackage{amsmath}
\usepackage{amssymb}
\usepackage{float}

\title{Three-dimensional micro-billiard lasers: the square pyramid.}

\author{M.~A.~Guidry\inst{1} \and Y. Song\inst{2}  \and C. Lafargue\inst{1} \and R. Sobczyk\inst{1}  \and D. Decanini\inst{3} \and S. Bittner\inst{1} \and B. Dietz\inst{2} \and L. Huang\inst{2}\and J. Zyss\inst{1} \and A. Grigis\inst{4} \and M. Lebental\inst{1}}
\shortauthor{M. A. Guidry \etal}

\institute{
\inst{1} Laboratoire de Photonique Quantique et Mol{\'e}culaire, UMR 8537, Ecole Normale Sup{\'e}rieure de Paris-Saclay, CentraleSup{\'e}lec, CNRS, Universit{\'e} Paris-Saclay, 94235 Cachan, France.\\
  \inst{2} School of Physical Science and Technology, and Key Laboratory for Magnetism and Magnetic Materials of MOE, Lanzhou University, Lanzhou, Gansu 730000, China.\\
  \inst{3} Centre de Nanosciences et de Nanotechnologies, CNRS, Universit\'e Paris-Sud, Universit\'e Paris-Saclay, C2N Marcoussis, 91460 Marcoussis, France.\\
  \inst{4} Laboratoire d'Analyse, G\'eom\'etrie et Applications, CNRS UMR 7539, Universit\'e Sorbonne Paris Cit\'e, Universit\'e Paris 13, Institut Galil\'ee, 99 avenue Jean-Baptiste Cl\'ement, 93430 Villetaneuse, France.
}

\abstract{Microlasers are of ample interest for advancing quantum chaos studies at the intersection of wave dynamics and geometric optics in resonators.  However,
the mode structures of three-dimensional microlasers without rotational symmetry remain largely unexplored due to fabrication limitations.
Previous studies of such cavities revealed lasing modes  localized on periodic
orbits exclusively confined to a single plane. In this work, we report on the characterization of pyramidal, polymer-based microlasers and demonstrate that the lasing modes are localized on a genuine three-dimensional periodic orbit. The consequences on the laser features are further discussed, in particular stability and polarization issues.}

\begin{document}

\maketitle

\section{Introduction}

Recent advances in the precise fabrication of three-dimensional (3D) microcavities open the door for original devices with innovative characteristics.
Direct Laser Writing (DLW)  allows for the fabrication of optical microresonators with sub-wavelength surface roughness \cite{2PP}. Thanks to this technology, we may now investigate long-standing questions like the influence of torsion on a wave, the nature and evolution of the polarization states and the sensitivity of the modes from potentially richer 3D geometry as compared to the two-dimensional (2D) case.

Among 3D geometries, optical resonators with rotational symmetry, such as spheres, ovoids, and toroids, have been extensively investigated and have shown great promise for producing highly confined whispering gallery modes \cite{tore-vahala}. However, such cavities are typically fabricated via melting techniques, greatly limiting the geometries which may be explored \cite{bouteille-2009,bouteille-2017,bouteille-chormaic}. The technology for precise control of semiconductor materials is nascent and the fabrication process arduous \cite{pyramid-1998,pyramid-karlsruhe,fabrication-hetterich,pyramide-APL}. However, the natural flexibility of polymer fabrication allows for the exploration of complex geometries. Among the many existing nanolithography methods, 3D two-photon lithography \cite{2PP}, based on a nonlinear optical effect allowing for local polymerization of a suitable resin, offers the unique possibility of creating almost arbitrary microstructures with controlled sub-wavelength precision in three dimensions.

Previous studies of cylindrical 3D cavities such as cuboids \cite{OE-3D,APL-FP3D} revealed lasing modes localized on periodic orbits. A \emph{periodic orbit} is a ray trajectory consisting of straight segments with specular reflections at the interfaces which returns to its origin after a certain number of bounces. In general, two-dimensional and three-dimensional billiards possess numerous periodic orbits\footnote{For instance, in a chaotic stadium billiard the number of periodic orbits with  length $L$ increases as $e^L$ \cite{brack}, whereas it is not clear if there exists at least one periodic orbit in an obtuse triangle \cite{triangle}.} ; such orbits play a central role in the semiclassical analysis of optical resonators and wave systems in general \cite{brack,stoeckmann-livre,plan-concave}. The periodic orbits found in Refs.~\cite{OE-3D,APL-FP3D}, however, were only two-dimensional in the sense that an individual orbit was confined to a single plane. In the present Letter, we focus on pyramidal microlasers, where the geometry discourages lasing on plane orbits. Instead, more complicated periodic orbits may emerge \cite{pyramide-orbites}, leading to torsion of the modes and non-trivial polarization states.

\begin{figure}
\onefigure[width = 0.7\linewidth]{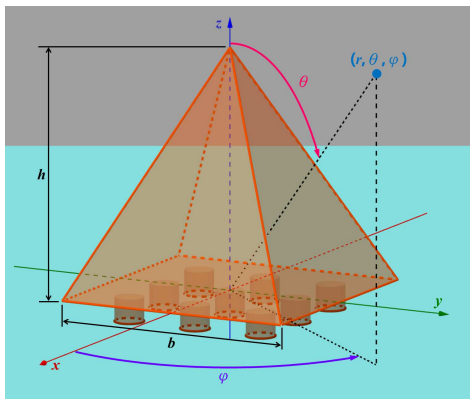}
	\caption{Schematic of the pyramid with side length $b$ and height $h=b$. The usual spherical coordinates ($r$,$\theta$,$\varphi$) are indicated.}
\label{fig:schema}
\end{figure}

\begin{figure}
\onefigure[width = 1\linewidth]{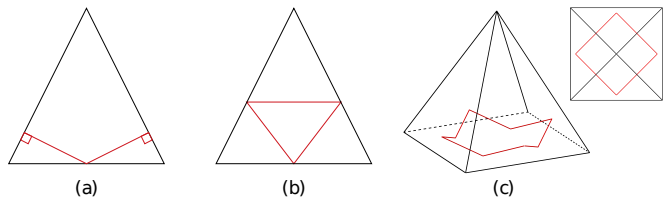}
	\caption{Examples of possible periodic orbits in the pyramid. (a) and (b) are plane orbits in the vertical ($yOz$) cross section. (c) The folded diamond periodic orbit. Inset: Top view of the folded diamond periodic orbit. The geometrical length of each trajectory is (a) $L=80~\mu$m, (b) $L=89.4~\mu$m, and (c) $L=133.3~\mu$m.}
\label{fig:periodic-orbits}
\end{figure}

In previous studies using 2D microlasers, we developed a robust method to infer the periodic orbit parameters from the experimental data \cite{plan-concave}. Over the course of several studies, we verified its validity in simple and well-controlled 3D geometries, namely cubes \cite{OE-3D} and 3D Fabry-Perot resonators \cite{APL-FP3D}. In the present study, we address the more complex case of polyhedra, in particular the pyramid with a square base shown in Fig. \ref{fig:schema}. Here the side length of the square base is $b=50\,\mu$m and equals the height $h$ ($\rho=b/h=1$); then the lasing modes localize on a non-planar periodic orbit (see Fig. \ref{fig:periodic-orbits}c). Since it looks like a diamond when projected onto the base (inset of Fig.\ref{fig:periodic-orbits}c), we refer to it as the \emph{folded diamond} periodic orbit. Indeed, this orbit shares some specific properties with the planar diamond periodic orbit, which dominates the laser features in 2D organic square microlasers \cite{EPL-carre}. Specifically, for a refractive index of $n=1.5$, this orbit has an angle of incidence on the lateral pyramid faces at the critical angle, which entails major consequences on the emission properties and the polarization states.

The outline of this Letter is the following. First, the fabrication process and the experimental setup are described. Based on the spectrum and the emission features, we then evidence that the lasing modes are localized along the folded diamond periodic orbit. Next, the consequences on the emission diagram and on the polarization states are discussed. Finally, the properties of the folded diamond orbit are studied  based on the cornerstone geometric concept of the screw angle, to serve as the basis for generalization to all types of convex polyhedra.

\section{Fabrication and experimental setup}

The pyramidal cavities were fabricated by Direct Laser Writing (DLW) lithography using a Photonic Professional GT system on the negative resist IP-G780 developed and manufactured by the Nanoscribe company. The resist was doped by $1$ wt\% pyrromethene 597 laser dye, provided by the Exciton company. We optimized the writing of the cavities which were exposed in a single step, layer-by-layer.  Applying DLW with the steering mirror technology, the writing time of a single cavity was about three minutes and the laser dye was not bleached. The unpolymerized photoresist was then removed using a developer\footnote{Developer: Propylene Glycol Monomethyl Ether Acetate.}, and rinsed with isopropanol.

\begin{figure}
\onefigure[width = 1\linewidth]{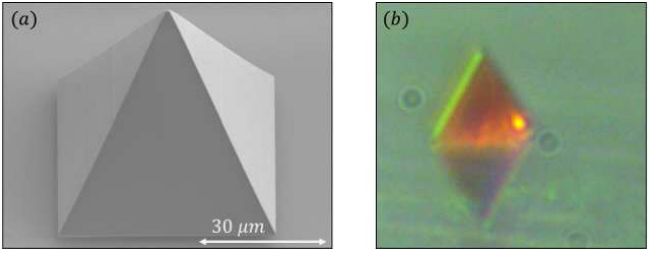}
\caption{(a) SEM image of a pyramid on posts. (b) Camera image of a pumped, lasing pyramid illuminated with white light. The lasing emission is the yellow hotspot on the right-hand side. The pump beam is not visible. The pyramid image is duplicated by reflection on the substrate.}
\label{fig:SEM-photos}
\end{figure}

The cavities were fabricated on nine $5$ $\mu$m-tall posts to avoid coupling with the glass substrate \cite{APL-FP3D}. A Scanning Electronic Microscope (SEM) image of a pyramid microlaser is shown in Fig.~\ref{fig:SEM-photos}a. In practice, the precision on the fabricated side lengths and on the SEM measurements is about $1$ $\mu$m. Since this implies a change of the periodic orbit length only in the third digit, this correction is not considered hereafter.\\

A single pyramid microlaser is pumped from above, perpendicular to the substrate, by a frequency-doubled pulsed Nd:YAG laser ($532$~nm, $500$~ps, $10$~Hz). The pump spot is much larger than the cavity to ensure uniform pumping. The emission features are not sensitive to the polarization of the pump beam. The experiments are carried out in air and at room temperature. The emission from the microlaser is collected by an optical fiber connected to a spectrometer and a cooled CCD camera with an overall spectral resolution of $0.03$~nm. The polar and azimuthal angles ($\theta$,$\varphi$) of the position of the collection fiber are controlled by a 3D goniometer \cite{gonio}. 

\begin{figure}
\onefigure[width = 1\linewidth]{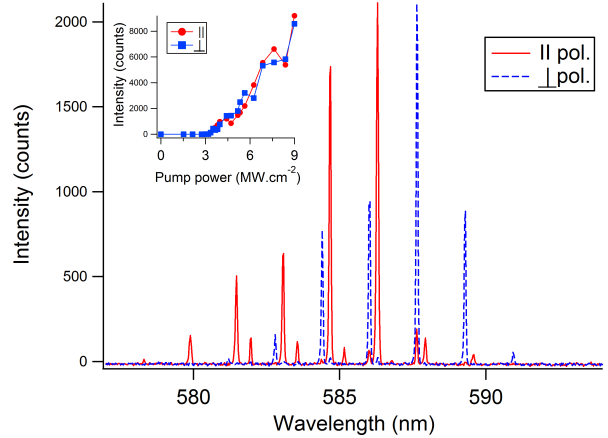}
\caption{Typical laser spectrum from a pyramid microlaser for two separate polarizations: the continuous red line corresponds to a polarization parallel to the substrate, and the dotted blue line polarization orthogonal to the substrate. Inset: Threshold curves for peaks at 585.56 nm and 585.88 nm  corresponding to both polarization states.}
\label{fig:spectre}
\end{figure}

\section{Spectrum and periodic orbits}

\begin{figure}
\onefigure[width = 1\linewidth]{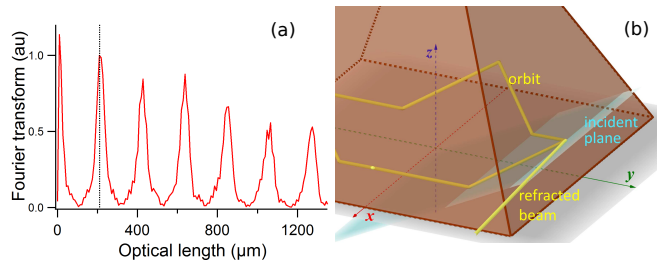}
\caption{(a) Normalized Fourier transform of the spectrum in Fig.\ref{fig:spectre}. The vertical dotted line indicates the expected position of the optical length for the folded diamond periodic orbit. (b) Scheme of the out-coupling by refraction at the lateral sidewall.}
\label{fig:FT}
\end{figure}

A typical laser spectrum is shown in Fig.~\ref{fig:spectre}. It consists of two slightly shifted combs of crossed polarizations, parallel and orthogonal to the substrate. This feature will be further discussed below. For each comb, the lasing peaks are almost equally spaced, which corresponds to longitudinal modes propagating along the same periodic orbit. Hence the wavenumbers fulfill a phase loop condition, leading to the following relation:
$$knL+\phi=2\pi\, m\hspace{1cm}m\in\mathds{N}$$
where $n$ is the refractive index, $L$ is the geometrical length of the periodic orbit, and $\phi$ is the global phase shift due to reflections after one round trip along the periodic orbit. As the wavenumbers are regularly spaced, the Fourier transform of the spectrum (as a function of $k$) features a single peak and its harmonics (see Fig.~\ref{fig:FT}a). The position of the first peak roughly corresponds to the optical length of the periodic orbit $nL$. In \cite{APL-FP3D}, corrections were considered in detail and it was shown that (i) the first peak of the Fourier transform is located at $n_gL$ where $n_g$ is the group refractive index, which includes dispersion, and (ii) an appropriate value of $n_g$ can be extracted from lasing spectra of cubes fabricated on the same sample. Here, measurements on cubes yielded $n_g=1.58\pm 0.02$.\\

In Fig.~\ref{fig:FT}a, the Fourier transform is peaked at $n_gL_{exp}=214\pm13~\mu$m, and this value does not vary significantly for different cavities or different directions of observation ($\theta$,$\varphi$). This implies an experimental geometrical length of $L_{exp}=135\pm9~\mu$m. The theoretical length of the folded diamond periodic orbit is $L=133~\mu$m, which is in excellent agreement with experiments. In Ref. \cite{pyramide-orbites} spectra were measured  from pyramid microlasers with a hexagonal base. The resulting Free Spectral Range (FSR) is consistent with a periodic orbit which is similar to the folded diamond periodic orbit. Other periodic orbits are excluded either because they are much shorter, such as the plane orbits depicted in Figs.~\ref{fig:periodic-orbits} (a) and (b), or because they are much longer, such as the diamond periodic orbit in the 2D square base of the pyramid \cite{octaedres}.\\

\section{Directions of emission}
The localization of modes along the folded diamond periodic orbit is also evidenced by the emission features. Figure~\ref{fig:SEM-photos}(b) shows an image of a lasing pyramid acquired by a camera with a high-magnification zoom lens. The lasing light is emitted from a bright spot on the pyramid face, close to the base, which is consistent with the position where the folded diamond orbit hits the pyramid face. Moreover, this image is observed at an angle nearly parallel to the pyramid base edge, suggesting that the emission stems from rays impinging on the cavity side walls with an angle of incidence very close to the critical angle, $\chi_{crit} = 41 \pm 1^\circ$, 
which approximately coincides with the angle of incidence of $41^\circ$ of the folded diamond periodic orbit on the lateral faces, see Fig. \ref{fig:FT}(b). Due to the uncertainties on the refractive index $n$ and on the side lengths of the cavity, it is impossible to find out whether the trajectory indeed undergoes total internal reflection (TIR) or is refracted. In either case, the reflection coefficient is close to unity and the emission is at grazing angle, as observed experimentally.

\begin{figure}
\onefigure[width = 0.6\linewidth]{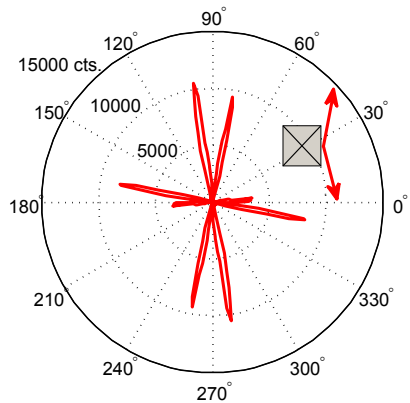}
\caption{Pyramid emission intensity with respect to $\varphi$, measured at $\theta = 85^\circ$. Due to technical limitations, the accessible angles were limited to $[-15^\circ, 195^\circ]$; the full
polar plot was created by duplicating the data. Insert: Scheme of the pyramid from the top, indicating the orientation. }
\label{fig:emission}
\end{figure}

\begin{figure}
\onefigure[width = 1\linewidth]{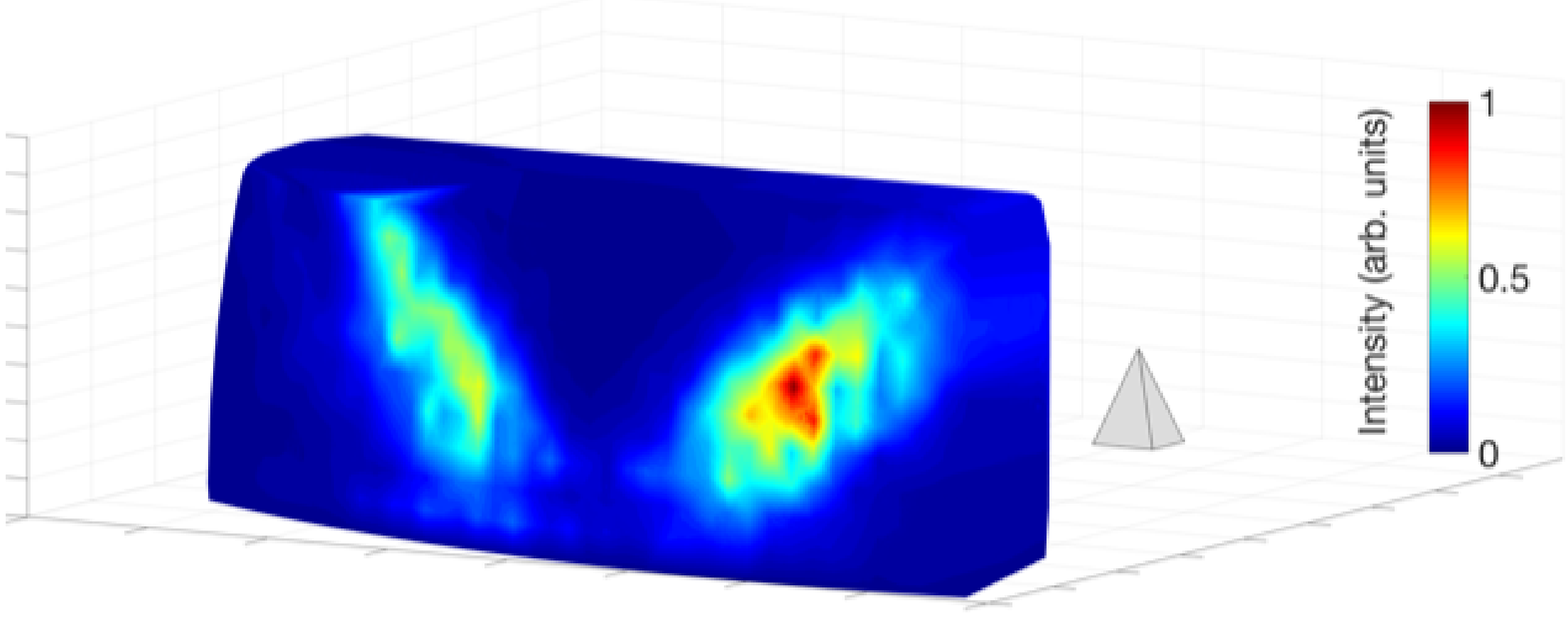}
\caption{Emission intensity with respect to $\theta\in [75^\circ, 90^\circ]$, and $\varphi \in [60^\circ, 120^\circ]$. }
\label{fig:3D}
\end{figure}

A typical far-field intensity distribution of a single resonance measured close to the substrate plane ($\theta=85^{\circ}$) is presented in Fig.~\ref{fig:emission}. The pattern exhibits eight emission lobes emitted nearly parallel to the side walls. Measuring from one face of the cavity, lobes are separated from the face normal by $\pm 10$ degrees. The relative intensity of the lobes slightly varies between cavities, and it is sensitive to the exact position of the pump spot, which seems to be  a specific feature of 3D microlasers compared to 2D microlasers.

A typical far-field intensity measurement of two emission lobes with respect to ($\theta$,$\varphi$) is presented in Fig.~\ref{fig:3D}. This out-of-plane measurement reveals that the two lobes rise out of the substrate plane along planes perpendicular to each respective pyramid face, forming a V-shaped emission.
The measured lobes appear refractive, where the region of maximal intensity corresponds to an angle of incidence just a fraction of a degree below the critical angle \cite{PRL-stone-2002}, consistent with the folded diamond orbit. This localization of modes along trajectories close to the critical angle was also noticed for the 2D organic square microlaser \cite{EPL-carre}: the quantization model showed that many modes could lead to lasing, whereas we observed only the modes which correspond to trajectories with reflections very close to the critical angle. The emission for trajectories confined by total internal reflection would be very low and hence not observable, whereas refractive escape for an angle of incidence well below the critical angle would lead to high losses, preventing lasing. Trajectories near the critical angle apparently provide a good compromise between strong, observable emission and low lasing threshold due to good confinement.

\section{Polarization of emission}
The emission near the critical angle directly
influences the mode polarization. As shown in Fig.~\ref{fig:spectre}, the lasing modes split into two orthogonal linear polarization states, parallel and orthogonal to the substrate, revealed by rotating a linear polarizer in front of the collection lens. The distinct mode combs have nearly equal intensity and laser threshold (see inset in Fig.\ref{fig:spectre}). It is not surprising to have two slightly shifted mode combs, as the phase conditions for both polarizations are different. However, it is not obvious that the two combs should have equal intensities and equal thresholds.

When a polarized plane wave undergoes total internal reflection (TIR) at an interface, the power is redistributed into another polarization state according to the angle of incidence and the plane of incidence. Moreover, the phase factor depends on the $s$ and $p$ Fresnel reflection coefficients. Hence, if light begins propagating through the cavity in a linearly polarized state, in general, it will not remain linearly polarized but rather turn into an elliptically polarized state.

The folded diamond orbit encounters eight reflections on the sidewalls. The incident angle of the four reflections on the base are at $\chi_b=70.5^{\circ}$, which is clearly above the critical angle $\chi_c=41^{\circ}$. The corresponding Fresnel reflection coefficients for $s$ and $p$ polarizations are then pure phase shifts, with unit modulus. On the contrary, the Fresnel coefficients on the lateral faces can be either real numbers less than one for an incident angle $\chi < \chi_c$, or pure phase shifts for $\chi>\chi_c$. In both cases,  $s$ and $p$ polarizations will be mixed because the incident planes of reflection at the base and at the lateral faces are not parallel, leading to elliptical polarization states. The angle $\chi=\chi_c$ is unique, because the Fresnel coefficients are unity. And there is no mixing of polarizations, since the folded diamond orbit is also highly symmetric. The eigenpolarizations hence remain the $s$ and $p$ polarization states relative to the base sidewall, which is precisely what is observed experimentally. \\
For a $\rho = 1$ pyramid with $n=1.5$, the folded diamond is just at the limit of the TIR condition. Thus the index of refraction preparing the orbit to be just on the cusp of TIR seems to play  a crucial role in both its prominence and its polarization-preserving nature.

\section{Stability}

\begin{figure}
\onefigure[width = 1\linewidth]{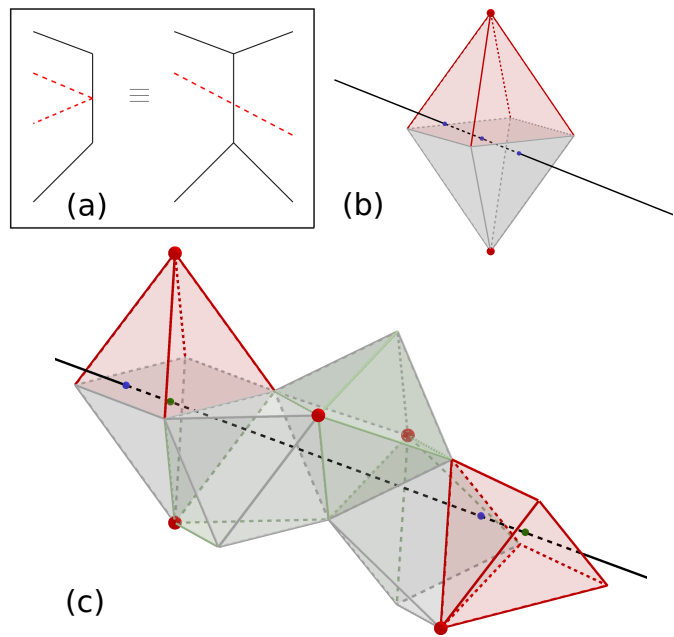}
\caption{(a) Unfolding procedure for a 2D cavity. The cavity boundary is drawn as a black line and the optical ray is the dotted red line. In the left-hand scheme, the ray is reflected, as usual. In the right-hand scheme, the cavity is reflected and the ray propagates in a straight line. (b) Unfolding of the pyramid for one reflection on the base. (c) Complete unfolding along the folded diamond periodic orbit. The screw angle corresponds to the rotation between the initial and final pyramids which are highlighted in red.}
\label{fig:screwing}
\end{figure}

The major difference between 2D and 3D euclidean geometries is that the rotations do not commute in 3D. The stability of periodic orbits is directly affected by this fundamental property and is quantified by the so-called \emph{screw angle}.\\
The screw angle can be defined by the unfolding procedure, which is described in
Fig.~\ref{fig:screwing} \cite{APL-1999-depliement}. When a ray is reflected on the wall, either the ray or the cavity can be reflected. In the latter case, the ray continues in a straight line. This reflection process is repeated for as many times as the trajectory reflects within the cavity boundary. In the end, the original polyhedron has undergone some global translation and rotation around the screw axis. This angle of rotation $\gamma$ is the screw angle.\\
Three different cases can appear: (i) If $\gamma$ is an integer multiple of $2\pi$, then the final polyhedron has exactly the same orientation as the original polyhedron. Hence, all trajectories which are parallel to the original periodic orbit are also periodic. (ii) If $\gamma$ is a rational multiple of $2\pi$ (i.e., $\gamma=2\pi\,p/q$, with $p,q\in\mathbb{N}$), then the rays which are parallel to the original periodic orbit are also periodic but with a period which is $q$ times longer. (iii) In the general case, $\gamma$ is not a rational multiple of $\pi$ and the original periodic orbit is isolated, namely there is no  periodic orbit in its vicinity. It is also \textit{stable} in the sense that it remains periodic when the polyhedron is subjected to small perturbations of the faces \cite{stabilite}.\\

The screw angle $\gamma$ only depends on the shape of the pyramidal cavity. For the pyramid with $\rho=1$ considered in this Letter\footnote{Detailed calculation of the screw angle by two different methods are provided in \cite{SPIE-2019}.},
\begin{equation}
    \gamma = 8[\pi - \arctan(3)]
\end{equation}
which is not a rational multiple of $\pi$. Hence, the folded diamond orbit is isolated and stable. However, $\gamma/2\pi\simeq0.4097$ (mod 1) is close to 2/5. So there exist nearby trajectories which are almost periodic, but with a $5$ times longer period than the folded diamond orbit.\\

According to the trace formula \cite{plan-concave}, the dominance of a periodic orbit in the emission spectrum is dictated by its round-trip losses  and its stability, compared to other periodic orbits. The theory remains to be developed in a 3D geometry with dielectric boundary conditions. However, we can extrapolate from existing theories, namely for 2D geometries with dielectric boundary conditions \cite{PRE-trace} and 3D geometries with metallic boundary conditions \cite{Balian-Duplantier}. Accordingly, we expect that the lasing modes preferentially localize on stable and well-confined orbits.  The periodic orbit shown in Fig.~\ref{fig:periodic-orbits}(a) is unstable with strong refractive losses, while the periodic orbit in Fig.~\ref{fig:periodic-orbits}(b) is stable, but with relatively high refractive losses. Hence the folded diamond orbit may dominate the spectrum, since it is stable and undergoes total internal reflection, consistent with the experiments.\\

\section{Conclusion}
In this Letter, we have evidenced that the lasing modes of a $\rho=1$ pyramidal microlaser with a square base are clearly localized on a specific periodic orbit, called the folded diamond orbit. Even though this orbit is not confined to a single plane, the polarization states are not elliptical due to the reflections on the sidewalls at the critical angle, or close to it. This specific property seems to point at the best trade-off for a laser between good refractive out-coupling  at the price of high losses, and good quality factor due to internal reflection but low emission, as it was already observed for 2D square microlasers \cite{EPL-carre}.

Pyramids exemplify the importance of torsion (e.g. departure of rays from planarity) and polarization in 3D cavities. The emergence of a dominant periodic orbit is  therefore likely to be controlled by such parameters as length, refractive losses, polarization state, and screw angle. For instance, in 3D metallic cavities, the contributions of periodic orbits with an odd numbers of reflections are vanishingly small \cite{Balian-Duplantier,3D-Darmstadt}. The trace formula, and more generally semi-classical physics  \cite{brack}, provides the framework to account for such properties. This has been so far developed only for 2D systems with dielectric boundary conditions \cite{PRE-trace}, while 3D ones remain an open frontier to be explored in the coming years. Fabrication techniques as well as characterization methods are now available and will provide the experimental counterpart for parallel theoretical studies. In this field, one can expect to harvest a wealth of original and surprising properties for various microlaser shapes, including knots, twists, m\"obius strips.

\acknowledgments
This work was partly supported by the French RENATECH network, the CNano IdF DIM Nano-K, the Labex NanoSaclay (ANR-10-LABX-0035), and the Laboratoire International Associ\'e ImagiNano. M.~A.~G. was supported by the  Franco-American Fulbright Commission. Y.S., B. D. and L. H. are supported by the National Natural Science Foundation of China under Grants No. 11775100 and No. 11775101. F. Di Mambro is acknowledged for the 3D paperboard patterns. The 3D drawings were made with Geogebra.

\end{document}